# Chipless Dielectric Constant Sensor for Structural Health Testing

# Chipless Dielectric Constant Sensor for Structural Health Testing

Antonio Lazaro, *Senior Member, IEEE*, Ramon Villarino, Filippo Costa, *Member IEEE,* Simone Genovesi, *Member IEEE*, Antonio Gentile, Luca Buoncristiani, David Girbau, *Senior Member, IEEE*

*Abstract*— A low-cost chipless RFID sensor tag to determine the dielectric properties of materials is presented. The sensor is based on a depolarizing dipole resonator loaded with printed capacitors whose capacitance depends on the permittivity of the material located on contact with the tag. The permittivity is obtained from the shift of the resonance frequency. The resonator is tested both individually as well as placed on a periodic arrangement so as to form a frequency selective structure (FSS) to increase the cross-polar radar cross section, thus facilitating the tag detection. The proposed structure and measurement procedure is particularly suitable for the characterization of civil materials as a non-destructive testing (NDT).

## I. Introduction

Chipless radio identification (RFID) tags are passive transponders that do not uses a microchip. Chipless RFID technology has emerged during last years in the literature. The main research efforts have been oriented to increase the number of bits that the electromagnetic structure can encode [1-3] or developing low cost printing techniques to manufacture low cost tags. However, this technology has also been proposed as a sensor [4-11]. In this case, due to the limited number of sensors within the read range of the tag, a large number of bits to perform the identification is not necessary. In this work, a chipless RFID tag is proposed as relative permittivity sensor that can be used for non-destructive testing (NDT). Non-destructive testing of materials is an important issue in civil structures [12], [13]. It is especially important in the case of concrete, since its structural properties depend on the mixture of compounds. The quality dependency of concrete in both the fresh and hardened states is dependent on many factors such as the type of cement, type of aggregates, water content, curing, and environmental conditions [13]. A wrong mixture could cause a structure to collapse. For these reasons, samples of the materials are periodically inspected in the laboratory. However, in situ methods and long-term non-destructive testing (NDT) of the civil structures are highly desirable. Chemical attack can stem from an ingress of aggressive agents (e.g. corrosion can follow carbonation or chloride ingress). Carbonation process is due to the chemical reaction of the dioxide from the environment and starts on the surface but progress to the interior. Carbonation results in degradation of concrete indirectly by reducing the pH. Due to carbonation, the permittivity decreases [14]. Microwave materials characterization techniques based upon dielectric property measurements have been proposed for detection and evaluation of physical and chemical changes in cement-based materials [14-16]. Several works have already proposed different wireless sensors for structural health monitoring [17-22]. A wireless sensor network is often used to collect the data from distributed sensors [12], [19]. A fiber optic link used for monitoring the strain of civil infrastructures at long distances is described in [17]. In the recent years, the RFID technology has been introduced for construction supply chain control [23]. A passive water-content sensor based on the detuning of a low-frequency resonator has been proposed in [18]. A similar concept is used for corrosion detection in [19] or dielectric permittivity measurements [20]. The detuning of the UHF RFID antenna is also used to sense dielectric constant in [21]. A UHF RFID integrated circuit that contains a temperature sensor has been proposed for monitoring the temperature during concrete maturation [24].

The tag consists of a Frequency-Selective-Surface (FSS). The unit cell element is a bent dipole over a ground plane loaded with capacitors whose capacitance depends on the relative permittivity of the material that is on contact with the tag. Therefore, its resonance frequency depends on the electrical characteristics of this material. The absence of a battery in the proposed tag makes it cheaper than the active ones and suitable for harsh environments. The paper is organized as follows. Section II presents the system and the sensing principle. Section III presents the simulations of the tag. Some preliminary experimental results are presented in Section IV. Section V presents a method to calibrate the tag and the experimental results. A low-cost reader based on Software Defined Radio (SDR) is presented in Section VI. Finally, Section VI draws the conclusions.

## II. Sensing principle

A chipless frequency-coded tag is based on one or several resonators that produces a change in the frequency response of

Manuscript received January 14th 2018.
This work was supported by the, H2020 Grant Agreement 645771–EMERGENT and the Spanish Government Project TEC2015-67883-R. The authors are with the Department of Electronics, Electrics and Automatic Control Engineering, Rovira i Virgili University, Tarragona, Av. Països Catalans 26, 43007 Tarragona, Spain. e-mail: antonioramon.lazaro@urv.cat. F. Costa, S.Genovesi, A. Gentile and L. Buoncristiani are with Department of Information Engineering, University of Pisa, Via G. Caruso, 56122 - Pisa, Italy. E-mail:filippo.costa@iet.unipi.it

the Radar Cross Section (RCS). Then, the detection is based on the measurement of the resonance frequency. However, the reflections at the environment (especially large objects near the tag) are often noticeably higher than the power reflected at the tag when conventional (non-depolarizing) resonators are used. This makes tags difficult to be detected in real environments, especially when attached to large elements, such as civil structures. To this end, this work uses of depolarizing tags [1], [2], [9-11]. Depolarizing tags based on resonators have been introduced in [2] and they show good performance when they are attached to lossy materials. Additionally, in this application, the reflections produced on the wall are mainly in the incoming polarization. Therefore, the use of a depolarizing tag overcomes this drawback.

A representation of the measurement system is shown in Fig.1. The tag is remotely read using a Vector Network Analyzer (VNA) that works as a reader. The reader sends a polarized wave (e.g. using a vertical oriented Vivaldi antenna) and then sweeps the interrogation frequency in order to measure the resonance frequency of the tag that depends on the material behind it (as will be explained later). The tag can be detected measuring the cross polar component using a cross-polarized antenna (e.g. using a horizontal oriented Vivaldi antenna) in reception. Therefore, the reception antenna filters the reflection on the wall and receives the backscattered field produced on the tag. In addition, the coupling interference between transmission and reception antennas is also minimized when using depolarizing tags.

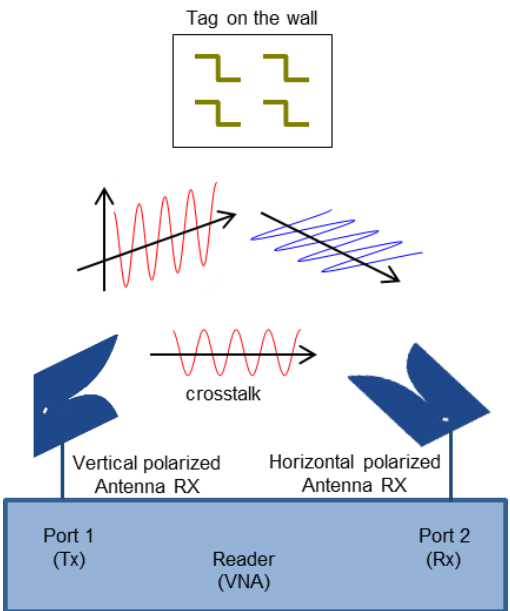

Fig.1. Schema of the measurement system.

A Frequency Selective Surface (FSS) composed of a depolarizing resonator (bent dipole) is used as dielectric constant sensor (Fig.2). It uses four-unit cell resonators to increase the level of the reflected signal and facilitate the sensor detection. The depolarizing resonator is based on a bent dipole loaded with two capacitors etched on the ground and connected to the resonant structure through vias. Fig.3. describes the layout of the unit cell. In Fig.3.b, the detail of the sensing capacitor formed by a slot ring around the small patch can be appreciated.

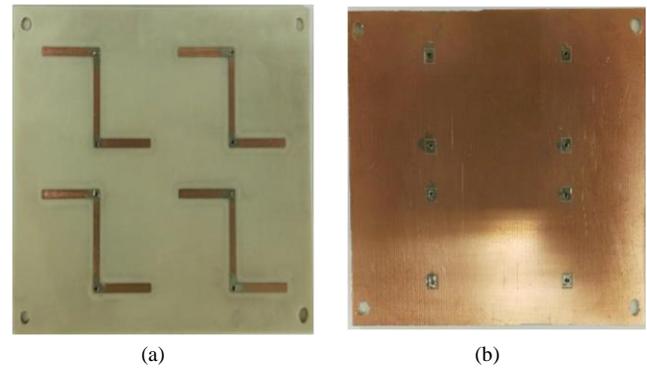

Fig.2. Prototype of the sensor tag composed by four unit cells. (a) Top view, (b) bottom view.

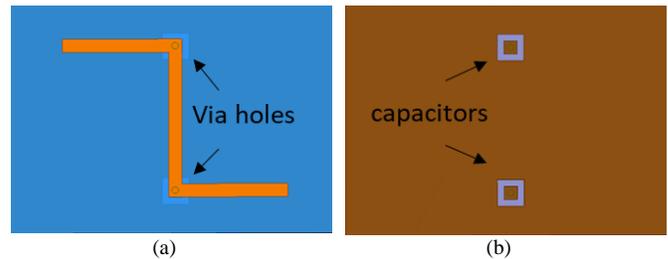

Fig. 3. Layout of the tag unit cell (a) top side, (b) bottom side. The via holes that connect the resonator in the top layer and the patch in the bottom layer have been marked. In the bottom, the sensing capacitor is formed by a slot ring around the small patch.

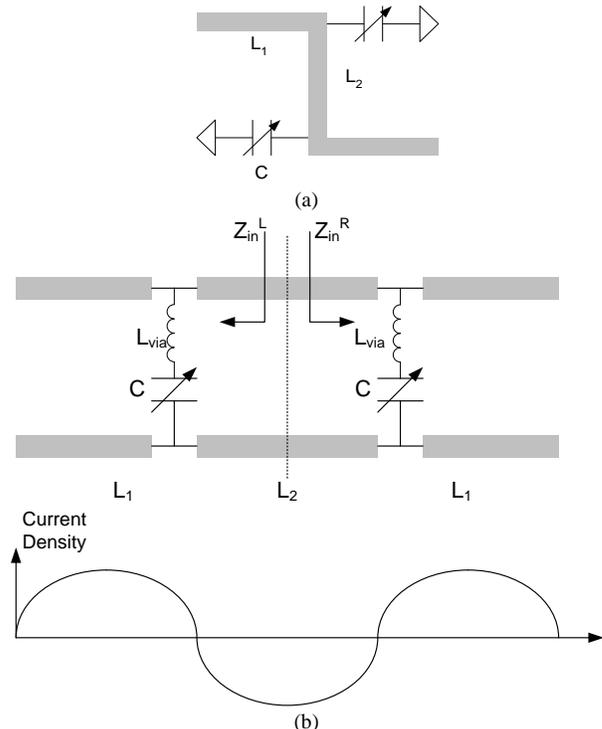

Fig. 4. Unit cell resonator (a), transmission line model and approximated current distribution (b).

In order to estimate the resonance frequency, a model of the resonator is shown in Fig.4.a where the distributed capacitor is modelled with its lumped equivalent. The perpendicular arms generate the depolarization of the incident wave and their length must be adequately tuned [10]. The resonance frequency can be tuned by changing the capacitor geometries. The capacitors are patterned in the ground plane in order to sense the materials attached to the tag (see the layout of Fig.3). Their capacitances depend on the slot width and the permittivity of the material. Therefore, a frequency shift in the resonance is expected as a function of the permittivity. A qualitative electrical model is shown in Fig.4.b. It consists of a section of microstrip lines with the two lumped capacitors. A simplified model for the capacitor is composed of the via inductance and the distributed capacitance.

## III. SIMULATIONS

In order to maximize the cross-polar component, the induced current in the lateral arms must be comparable to the current in the center of the dipole that contributes mainly to the co-polar component of the radar cross section. The current at the ends of the unit cell resonator must be zero imposed by the open circuit condition. The capacitor presents a low impedance at the resonance frequency. Therefore, a low value of current is expected at this point. An approximated current density distribution along the resonator is shown in Fig.4.b. This current distribution has been verified by means of electromagnetic simulations that will be shown below. It can be expected that the 3rd resonance increases the cross-polar components. Therefore, a simplified circuit of Fig.4.b for the determination of resonance frequency can be proposed. The equivalent circuit is composed by a microstrip transmission line of length $L_2$ that models the section of the resonator oriented in the copolar direction (vertical direction in Fig.1-2). This line is loaded with two microstrip transmission lines of length $L_1$ that model the propagation in the bent lines. The sensors at the corners are modelled as capacitors whose capacitance changes with the dielectric constant of the material under the ground plane. An inductor connected in series with the capacitance between the slot ring and the ground models the inductance of the via hole that connects the strip line printed on the top layer and the small patch situated just under the line on the bottom layer (see Fig.3).

The resonance frequency can be obtained from the transverse resonance method [25]. If the middle point in the model of Fig.4.b is considered, the impedance $Z_{in}^R$ looking forward must be equal to minus the impedance $Z_{in}^L$ looking backward, so that $Z_{in}^L + Z_{in}^R = 0$. The other resonances can be obtained applying the dual condition, $Y_{in}^L + Y_{in}^R = 0$.

The impedance from the middle point can be obtained from the input impedance of a transmission line of length $L_2/2$ and a characteristic impedance $Z_0$ loaded with an impedance $Z_L$:

$$Z_{in}^{L,R} = Z_0 \frac{Z_L + jZ_0 tan(\beta L_2/2)}{Z_0 + jZ_L tan(\beta L_2/2)} \quad (1)$$

Where $Z_L$ is the parallel combination impedance of the capacitance $C$ in series with inductance $L_{via}$ and the open circuit transmission line of length $L_1$:

$$Z_L = \frac{1}{\frac{1}{j\omega L_{via} + 1/(j\omega C)} + j\left(\frac{1}{Z_0}\right)tan(\beta L_1)} \quad (2)$$

Fig.5 shows the variation of the 3rd resonance frequency as function of the capacitance C. It is numerically obtained solving the resonance condition equation using a zero-search routine (with Matlab) and it is checked using a circuit simulator (Keysight ADS). An inductance of 2.5 nH for the via hole is considered. It is demonstrated that, for capacitances smaller than 2.5 pF, the behavior is nearly linear, therefore it is expected to obtain a linear dependence with the permittivity of the material located behind the ground plane. Fig.6 shows the simulation of the capacitance of the slot capacitor (C) performed with Keysight Momentum) with a gap of 0.8 mm as function of the dielectric permittivity of the material behind the tag. It can be observed that the capacitance C varies linearly with the permittivity.

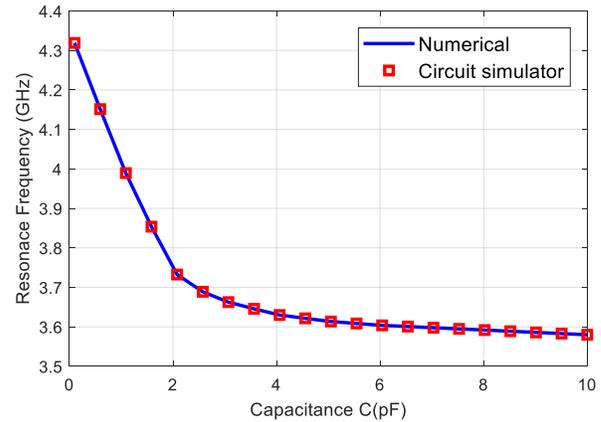
Fig.5. Resonance frequency as a function of load capacitance computed with the transmission line model (-) and the circuit simulator (□).

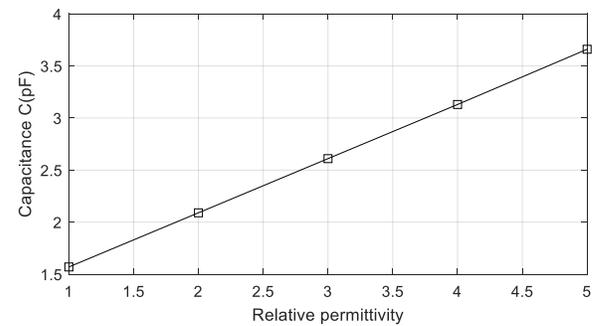
Fig.6. Simulated capacitance of the slot capacitor as function of the relative permittivity of the material.

As a proof of concept, a design at 4 GHz is performed. This frequency is selected because it matches with the peak gain of the antennas used in the experimental setup and where the maximum isolation is achieved. However, the design can be easily scaled to other frequencies by changing the length of the dipoles. The chosen lengths of dipole arms are $L_1$=20 mm, $L_2$=22 mm and the width is 2 mm. The via's hole diameter is 1 mm and the slot gap in the ground plane is 0.8 mm. The parameters have been chosen keeping a compromise between the quality factor of the resonance and the level of cross polar radar cross section. From the image electromagnetic theory, the image currents induced in the ground plane under the strips have opposite sign with respect the current in the strips. Therefore, a partial cancellation of the radiation fields is produced due to these currents. Small substrate thickness results on high quality factors but the cross polar RCS is smaller than when thicker substrates are used. The substrate used in these prototypes is Roger RO4003 with 64 mil thickness in order to facilitate tag detection. However, to derive a simple model from the slot capacitance is not straightforward. For this reason,

HFSS 3D electromagnetic simulations are used to confirm the previous model. Fig.7 shows the current distribution of the unit-cell element at the frequency that maximizes the cross-polar component. The current distribution confirms the behavior predicted in Fig.4b. The high current introduced in the folded arms produces the cross-polar component of the radar cross section.

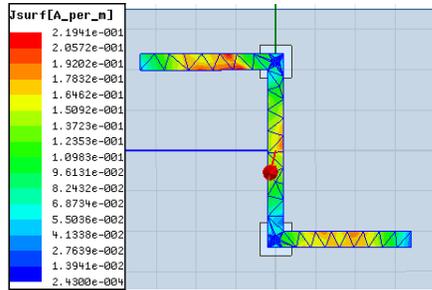

Fig. 7. Current density distribution in the resonator.

Fig.8 shows HFSS simulations of the cross-polar RCS for materials located behind the tag with different permittivities ($\varepsilon_r$). A smooth frequency shift, of the same order than the predicted by the transmission model, is appreciated. The width of the slot in the ground plane (see Fig.3) can be used to perform a fine tuning of the resonance frequency as it is shown in the simulations of Fig.9. In practice, this parameter is sensitive to the layout tolerance, especially if the printed board is manufactured using custom facilities. Fig.10 shows the effect of loss introduced by the material attached behind the tag. To avoid the dependence of the result with the thickness of material a 10 mm slab is considered in the simulations, since in real environments tags will be embedded on the surface of a thick material under test. The quality factor decreases, and a frequency shift is predicted. Therefore, the tag can be useful for materials with low or moderate losses, being wet samples (e.g. fresh concrete) very difficult to detect.

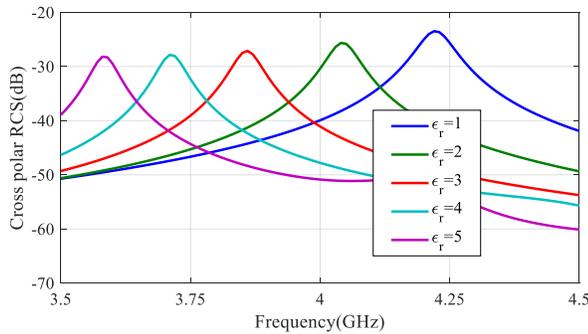

Fig.8. Simulated cross-polar RCS as a function of the frequency for different permittivity of the material behind.

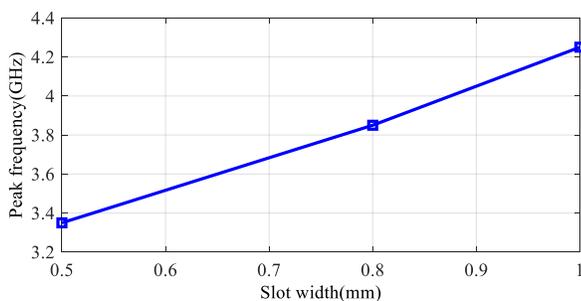

Fig. 9. Simulation of the tag response on air with respect to the slot width.

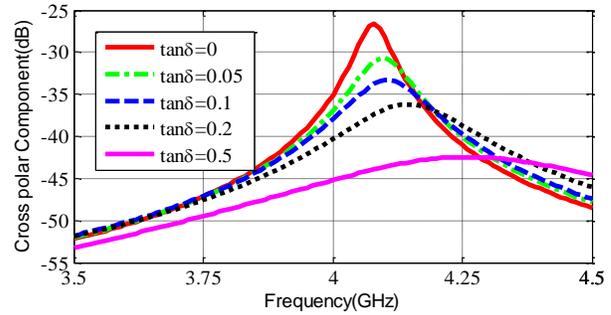

Fig. 10. Effect of material losses in the tag response.

## IV. PRELIMINARY EXPERIMENTAL VALIDATION WITH BACKGROUND SUBTRACTION

In order to increase the RCS and the detectability, the manufactured tag is composed by four unit cells, as shown in Fig.2. Fig.11 reports the measurements of backscattered cross-polarized fields for different materials with known permittivity. These results will be used to calibrate the sensor. The thickness of the materials behind is larger than 10 mm to avoid dependence of the frequency shift also on the thickness. The results are obtained using the background subtraction technique that consists of subtracting the measurement of the scene without the tag from the measured parameter $S_{21}$ in presence of the tag. The same smooth frequency shift observed in simulations are found. The resonant frequency obtained is slightly lower than in simulations due to the coupling between elements and the tolerances in the fabrication of the slot and the via holes using a custom circuit manufacturing facility. Therefore, the experimental permittivity calibration procedure may be required. Fig.12 shows the variation of the frequency peak obtained from Fig.11 as a function of the material permittivity. A linear regression is made with good agreement. This line can be used to measure the permittivity of unknown materials. An additional important parameter is the read range and the repeatability of the measurement when the distance changes. Fig.13a shows the measurement of the tag on the air as a function of the reader-to-tag distance. When the distance increases, the received level clearly decreases however the peak frequency remains constant. A read range close to 1 m is estimated, if a minimum signal-to-noise ratio of 10 dB is assumed. The read-range can be increased by adding more unit cells, at the expense of increasing the tag size as can be shown in Fig.13b. The resonance frequency of the tag with only one element (unit cell) and the FSS with 2 x 2 elements is the same but the range achieved is higher in this case compared to the one element case, 100 cm in front of 40 cm An improvement of 12 dB is obtained as it was expected due to increase of the radar cross section of the FSS that depends of $N^2$, where N is the number of cells [26].

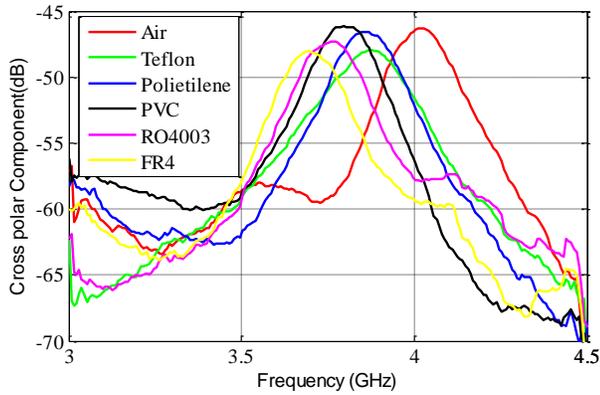

Fig. 11. Measured cross-polar component after background subtraction for different materials behind the tag.

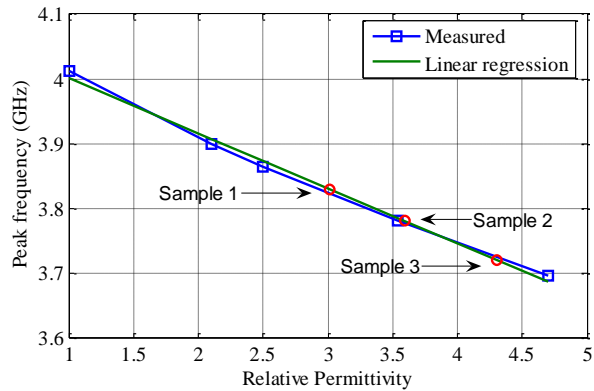

Fig.12. Measured peak frequency as a function of the permittivity of known materials (□). A linear regression is applied to measured resonances of sample materials (o).

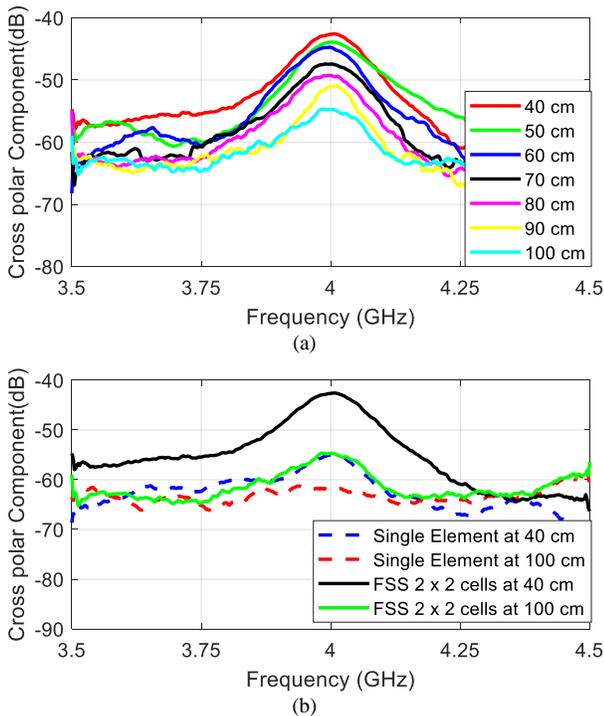

Fig. 13. (a) Measured response of the tag composed by 4 elements for different reader-to-tag distances. (b) Measured response of the tag composed by 4 elements and single element at distance of 40 cm and 100 cm

## V. ON-SITE CALIBRATION TECHNIQUES

### A. Calibration

One of the main difficulties in chipless RFID technology is the calibration needed for tag detection. Usually the background subtraction technique is used, but this approach is unpractical in most real applications, since it requires to remove the tag to measure the scene [1-3]. For long-term NDT of civil structures, the tag is inserted during the fabrication of the structure and it is periodically measured (after the sample is dry). Therefore, the tag cannot be removed to perform background subtraction. In order to solve this problem some alternatives have been investigated.

First of all, an analysis of the received signal (parameter $S_{21}$ measured between the transmit and receive antennas) is taken into account. The measured signal can be modelled as the sum of three contributions: the tag response, $s_{tag}(t)$, the coupling interference between the antennas, $s_c(t)$, and the clutter due to the backscattered field on the surrounding objects, $s_{clutter}(t)$.

$$s(t) = s_{tag}(t) + s_c(t) + s_{clutter}(t) \qquad (3)$$

The coupling signal is due to the crosstalk between the antennas and depends on the antennas used and the distance between them. It can be considered constant with respect to the distance between the tag and reader and often it presents a much higher amplitude than the tag and the clutter. The amplitude of the tag response depends on the distance between tag and reader. The clutter response can be split in a couple of terms: one due to the objects close to the tag and the other due to the objects far from the tag. The first contribution presents higher amplitude and it often strongly depends on frequency. In case of depolarizing tags, the reflection in flat surfaces such as walls or ground is minimized because co-polar component that is filtered by the receiver antenna in cross polarization is produced. Following the model (3), the tag response can be obtained subtracting a reference measurement without tag response ($s_{ref}(t) \approx s_c(t)+s_{clutter}(t)$) to the measurement of the tag. Therefore, the reference measurement can be obtained eliminating the tag response by locating an absorbent that attenuates the reflection in both polarizations or placing a metallic plate that eliminates the reflection in the cross polar component. In both cases, it is assumed that the clutter is invariant along the measurement with the tag and the calibration with the reference. The metallic plate can be replaced by the wall or another material where the tag is attached, assuming that the distance remains invariant.

Two different practical scenarios are considered. In the first one, the read range is very small (20-30 cm), and in the second one the distance to the material under test is higher (50 cm). In all the cases, the crosstalk between antennas dominates, especially outside the resonance frequency of the tag, but it is removed using any of the previous reference measurements. In the first case, the tag response dominates over the clutter contributions, whereas in the second the clutter amplitude is dominant. For small read range (Fig.14.a), the tag can be detected easily from the raw measurement without subtracting any reference measurement because the amplitude of the backscattered field in cross polarization is higher than the other contributions, mainly in copolar components. Therefore, only one measurement is needed. However, for longer distances (Fig.14.b) it is not possible to detect the tag from the raw measurement.

Experimental results with the tag attached on a wall of bricks in the laboratory (indoor environment) located at 20 cm and 50 cm from the antennas are shown in Figure 14.a and 14.b, respectively. Fig. 15 shows a photograph of the experiment where the tag is on contact with the wall. These figures show the raw measurement (parameter $S_{21}$ between two cross-polarized antennas), and the signal obtained by subtracting the following reference measurements: the wall without the tag, a metallic plate (used as reference tag), an absorbent material that covers the tag and the crosstalk measurement of the antennas. A piece of Eccosorb VHP-4 with a typical reflectivity of -30 dB at the operating band with the same area of the tag and thickness of 5 cm is used as absorber. In both of the tested scenarios, the tag can be detected using a metallic plate or an absorber as background, or subtracting the crosstalk between antennas. This crosstalk measurement can be stored once and subtracted to all tag measurements. A good agreement is obtained between the three calibration procedures. As conclusion, the detection of the proposed permittivity sensor based on a FSS made of depolarizing resonators can be measured up to 50 cm.

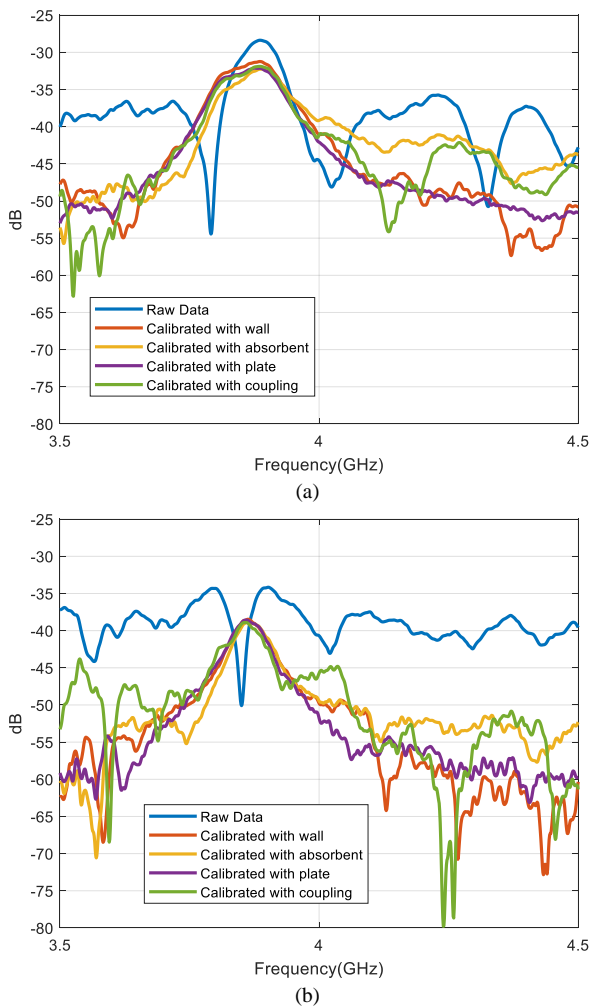

Fig.14. Comparison of the raw measurement of the tag on a wall at 20 cm (a) and 50 cm (b) for different calibration procedures.

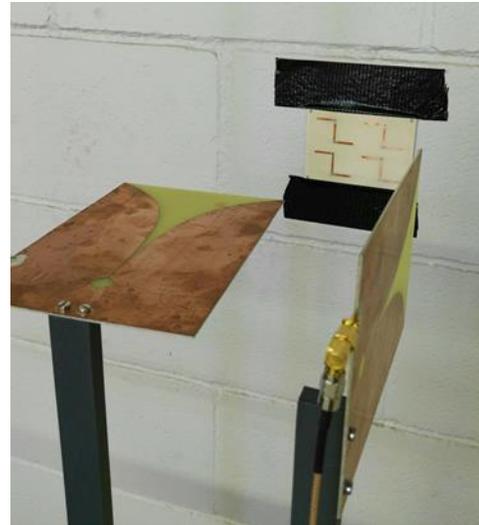

Fig.15. Photography of the experimental setup for measurement of the tag in a wall.

### B. Comparison with other dielectric measurement techniques

Finally, the measurement method is tested, in a realistic environment, with different types of walls. Sample 1 is a wall of concrete (100% concrete), sample 2 is a block of 50% concrete and 50% sand and sample 3 is a wall of bricks. In order to compare the values of permittivity obtained with the chipless sensor, the samples have also been measured with the ring resonator method [6]. In this characterization method, the permittivity is obtained from the frequency shift of the resonance frequency [27]. A ring resonator printed on 1.6 mm thick Rogers 4003 has been designed to resonate around 4 GHz (see Fig.16). The first resonance frequency can be obtained from the effective permittivity $\varepsilon_{ref}$ given by [28]:

$$f_r = \frac{c}{2\pi R \sqrt{\varepsilon_{ref}}} \quad (4)$$

$$\varepsilon_{ref} = \frac{\varepsilon_{rs}+\varepsilon_{rmat}}{2} + \frac{\varepsilon_{rs}-\varepsilon_{rmat}}{2}\left(1+\frac{12W_e}{h}\right)^{-1/2} \quad (5)$$

Where $R$ is the radius of the resonator (6.9 mm), $W$ is the width of the line (1.85 mm), $h$ is the thickness of the substrate (32 mil) and $t$ is the metallization thickness (34 μm). $\varepsilon_{rs}$ and $\varepsilon_{rmat}$ is the relative permittivity of the substrate and the material, respectively. $W_e$ is the effective width given by:

$$W_e = W + \frac{t}{\pi}\left(1 + ln\left(\frac{2h}{t}\right)\right) \quad (6)$$

When the material under test is located on the top of the resonator, a shift in the resonance frequency is produced due to the variation of the effective permittivity. Fig.17 shows the resonator's relative frequency shift, considering the presence and absence of a material on the top. The expressions (4)-(6) have been checked with electromagnetic simulations with Keysight Momentum. Fig.18. depicts an example of measurement, where both the frequency shift and the reduction of the quality factor due to the presence of the block of concrete are clearly visible. Table I compares the permittivity obtained using the calibration line obtained in Fig.12 and the measurement of the samples using the ring resonator method. The average difference between the two methods is 3.94%.

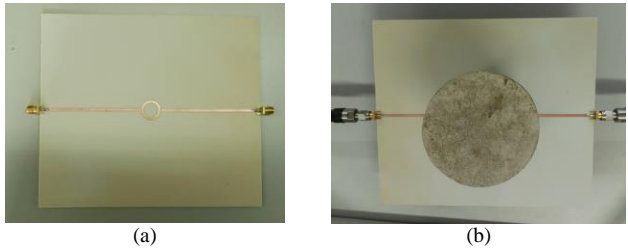

Fig.16. Photography of the unloaded ring resonator (a) and loaded with a block of concrete (b).

TABLE I
SAMPLE COMPOSITION AND PERMITTIVITY

| Sample | Material | Measured permittivity Chipless tag | Measured permittivity Ring resonator method | Difference (%) |
|---|---|---|---|---|
| 1 | Concrete 100% | 3.56 | 3.51 | 3.55 |
| 2 | Concrete 50% | 3.98 | 4.07 | 3.97 |
| 3 | Wall of bricks | 4.31 | 4.24 | 4.30 |

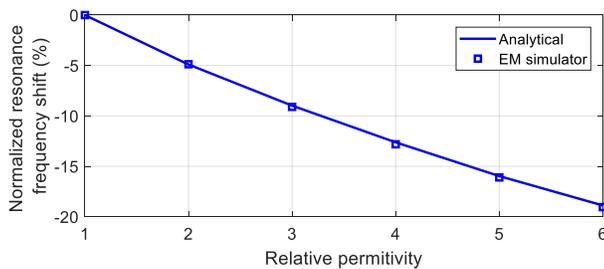

Fig.17. Normalized frequency shift with respect the ring resonator loaded with the air as function of the material relative permittivity. Comparison between the analytical (solid line) model and EM simulations (□) with Keysight Momentum.

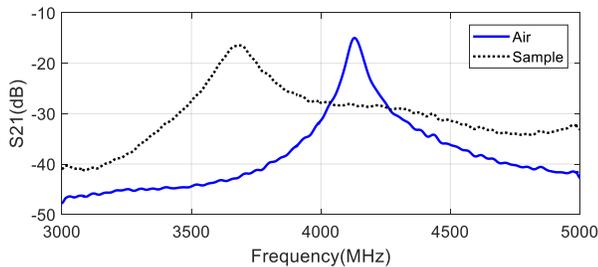

Fig.18. Measured $S_{21}$ for the ring resonator unloaded (in air, solid line) and with a sample block of concrete (dotted line).

### C. Effect of humidity and temperature

In order to study the influence of the moisture in the sample some experiments have been done. First, a dry concrete wall is measured. Later the wall has been wet with water and the measurement has been repeated. The results are shown in Fig.19. A shift in the measurement is depicted due to the moisture, but the peak frequency returns to the dry value when the sample tends to dry state after 1 h. The absorption of moisture by the substrate used to print the tag due to the ambient humidity can also affect the measure. In order to quantify it a ring resonator is introduced in a climatic chamber. Fig. 20 shows the variation of the permittivity of the Rogers 4003 substrate over a cycle from 0 to 100%. A variation of 0.1 in the permittivity and a recover of its initial value when substrate is dried can be observed. The reduction of the quality factor or increase of the tangent losses as consequence of the water absorption is also observed. The result of variation in the resonance frequency of the ring resonator is 1%. It can be expected that the variation of the frequency of the chipless tag due to ambient humidity will be of the same order.

The temperature can also affect the measurement of the resonance of the chipless tag. A small change in the permittivity of the substrate and the expansion of the resonator is produced. Fig. 21 compares the measurement of a tag (at approximately distance of 20 cm from the reader) on the air at 25ºC and after heating with a heat gun at 100ºC. The variation is -9.1 MHz. This variation is smaller than the bandwidth (about 100 MHz) of the measured peak. Therefore, the error can be neglected for the typical temperature measurement range.

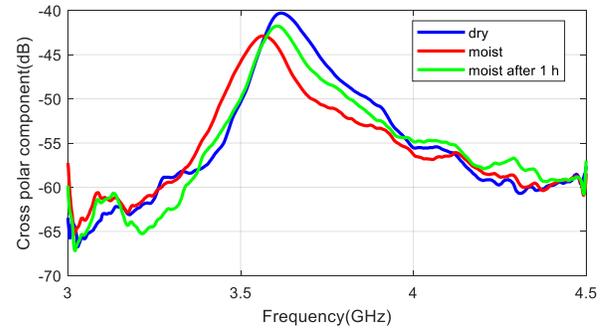

Fig.19. Comparison of a measurement of a wall of concrete as function of the moist degree.

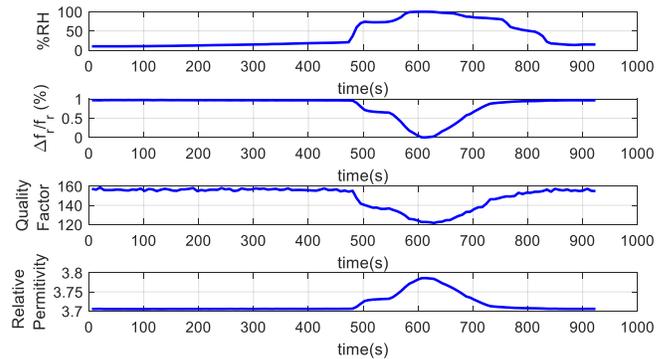

Fig.20. Variation of the resonance frequency, quality factor and relative permittivity of a ring resonator made with Rogers 4003 as function of the time when the relative humidity change (top).

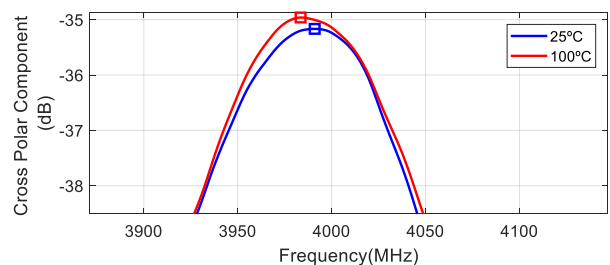

Fig.21. Comparison of a tag measurement at 25 ºC and 100 ºC.

## VI. Detection with Low-cost Software defined radio (SDR) reader

One important drawback of chipless technology over well-established commercial RFID technologies (HF or UHF RFID) is the lack of commercial readers. Therefore, the measurements are often performed using VNA available at laboratories. However, some proof of concept prototypes has been proposed in the literature. A prototype based on a VCO and a phase/gain detector (AD8302) between 1.9 GHz and 2.5 GHz is reported in [29]. Software defined radio (SDR) N210 USRP platform based on powerful FPGA is used as reader in [30][30]. Another reader based on time-domain UWB sampler is presented in [31], [32]. Additionally, low-cost VNA can be found in the market that can replace benchtop VNA for field measurements. Experimental results of section V.A shows that for small distances, scalar measurements can achieve good precision because the scattered field by the tag is high enough to neglect the coupling. Therefore, these results open the door to use simplified low-cost readers such as spectrum analyzers with tracking generator option or low-cost receivers as described next. As a proof of concept, a low-cost reader based on a popular low-cost SDR receiver (RTL-SDR based on RTL2832U chipset) is used. As the frequency band of the RTL2832U is up to 2 GHz, a down-converter based on a mixer (LRMS30J from Minicircuits) is used. The local oscillator for the mixer and the transmitter can be implemented using a phase-lock-loop (PLL) synthesizer (from IDM Instruments). The received signal at fixed IF (1.5 GHz) is sampled by the RTL-SDR receiver. The data is transferred to the computer by USB. After filtering the data, the Hilbert transform is applied to the sampled signal obtaining both amplitude and phase information. The reader is powered by USB, therefore a portable reader can be implemented. Fig.22 shows the block diagram and a photography of the prototype. Fig. 23 shows a short range (20 cm) typical tag measurement which achieves good agreement with that obtained with the VNA.

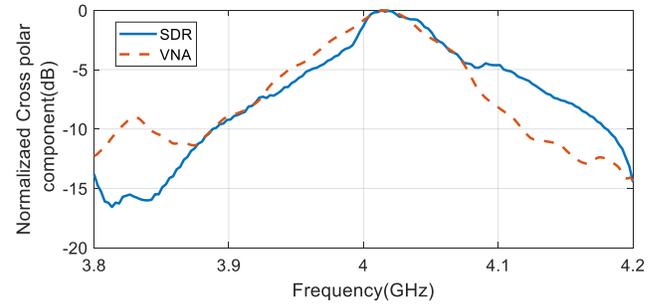

Fig.23. Comparison of a measurement as function of the frequency performed with the SDR based reader and with VNA.

## VII. Conclusion

A novel cross-polarized chipless sensor based on a depolarizing dipole loaded with capacitors is proposed as permittivity sensor for long-term non-destructive testing of structures. A simple model based on a loaded transmission line is presented. Simulated and experimental results show a nearly-linear dependence of the resonance frequency of the tag with the permittivity, which permits the calibration of the sensor. The tag has been tested in realistic scenarios and different procedures to detect the sensor on large civil structures are proposed, which do not require tag removal, as an alternative to background subtraction. For small distances the tag can be detected from the raw measurement without subtracting any background subtraction. However, for higher distances, the tag can be detected subtracting the crosstalk measurement or using a reference tag composed by a metallic plate in front the tag or an absorbent. The influence of temperature, moisture and ambient humidity is empirically analyzed. Finally, the feasibility of detecting the sensor using a low-cost reader based on SDR receiver has been demonstrated for short ranges.

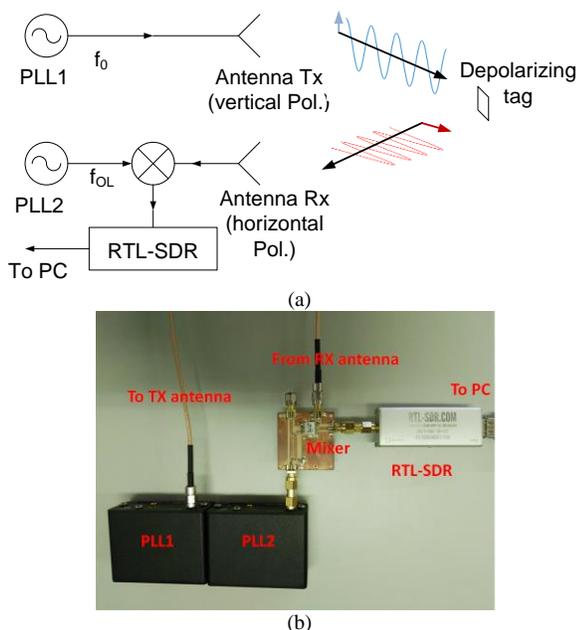

Fig.22. (a) Block diagram and photography (b) of the reader based on a low-cost SDR receiver.

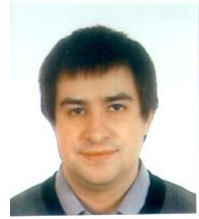

**Antonio Lázaro** (M'07, SM'16) was born in Lleida, Spain, in 1971. He received the M.S. and Ph.D. degrees in telecommunication engineering from the Universitat Politècnica de Catalunya (UPC), Barcelona, Spain, in 1994 and 1998, respectively. He then joined the faculty of UPC, where he currently teaches a course on microwave circuits and antennas. Since July 2004, he is a Full-Time Professor at the Department of Electronic Engineering, Universitat Rovira i Virgili (URV), Tarragona, Spain. His research interests are microwave device modeling, on-wafer noise measurements, monolithic microwave integrated circuits (MMICs), low phase noise oscillators, MEMS, RFID, UWB and microwave systems.

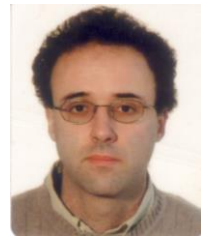

**Ramon Villarino** received the Telecommunications Technical Engineering degree from the Ramon Llull University (URL), Barcelona, Spain in 1994, the Senior Telecommunications Engineering degree from the Polytechnic University of Catalonia (UPC), Barcelona, Spain in 2000 and the PhD from the UPC in 2004. During 2005-2006, he was a Research Associate at the Technological Telecommunications Center of Catalonia (CTTC), Barcelona, Spain. He worked at the Autonomous University of Catalonia (UAB) from 2006 to 2008 as a Researcher and Assistant Professor. Since January 2009 he is a Full-Time Professor at Universitat Rovira i Virgili (URV). His research activities are oriented to radiometry, microwave devices and systems, based on UWB, RFIDs and frequency selective structures using MetaMaterials (MM).

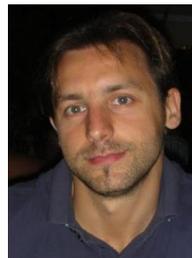

**Filippo Costa** (S'07-M'11) received the M.Sc. degree in telecommunication engineering and the Ph.D. degree in applied electromagnetism from the University of Pisa, Pisa, Italy, in 2006 and 2010, respectively. In 2009, he was a Visiting Researcher at the Department of Radio Science and Engineering, Helsinki University of Technology, TKK (now Aalto University), Finland.

He is currently an Assistant Professor at the University of Pisa. His research interests include metamaterials, metasurfaces, antennas and Radio Frequency Identification (RFID).

He serves as an Associate Editor of the IEEE SENSORS LETTERS. He was appointed as outstanding reviewer of IEEE TRANSACTIONS ON ANTENNAS AND PROPAGATION in 2015, 2016 and 2017 and IEEE ANTENNAS AND WIRELESS PROPAGATION LETTERS in 2017. He received three times the URSI Young Scientist Award in 2013, 2014 and 2015.

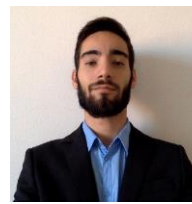

**Antonio Gentile** received Bachelor and M.Sc. degree in Telecommunications Engineering from the university of Pisa, Pisa, Italy, in 2013 and 2016, respectively. From August 2016, he is Research Assistant at the Microwave and Radiation Laboratory, Information Department, University of Pisa. From February 2017 to December 2017, he was a Visiting Researcher at University Rovira i Virgili, Tarragona, Spain, working on chipless RFID sensors.

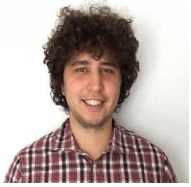

**Luca Buoncristiani** received Bachelor and M.Sc. degree in telecommunications engineering from the University of Pisa, Italy, in 2013 and 2016, respectively. In 2016, he was Research Assistant at the Microwave and Radiation Laboratory, Dipartimento dell'informazione, University of Pisa. During the period 2016-2017 he was a Visiting Researcher at University Rovira I Virgili, working on chipless RFID sensors. Currently he is with U-blox and he is involved in design and testing of cellular modules.

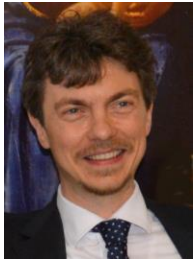

**Simone Genovesi** (S'99-M'07) received the Laurea degree in telecommunication engineering and the Ph.D. degree in information engineering from the University of Pisa, Pisa, Italy, in 2003 and 2007, respectively. He is currently an Assistant Professor at the Microwave and Radiation Laboratory, University of Pisa. Current research topics focus on metamaterials, radio frequency identification (RFID) systems, optimization algorithms and reconfigurable antennas. He was the recipient of a grant from the Massachusetts Institute of Technology in the framework of the MIT International Science and Technology Initiatives (MISTI).

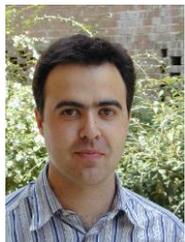

**David Girbau** (M'04, SM'13) received the BS in Telecommunication Engineering, MS in Electronics Engineering and PhD in Telecommunication from Universitat Politècnica de Catalunya (UPC), Barcelona, Spain, in 1998, 2002 and 2006, respectively. From February 2001 to September 2007 he was a Research Assistant with the UPC. From September 2005 to September 2007 he was a Part-Time Assistant Professor with the Universitat Autònoma de Barcelona (UAB). Since October 2007 he is a Full-Time Professor at Universitat Rovira i Virgili (URV). His research interests include microwave devices and systems, with emphasis on UWB, RFIDs, RF-MEMS and wireless sensors.